# Ligand-Surface Interactions and Surface Oxidation of Colloidal PbSe Quantum Dots Revealed by Thin-film Positron Annihilation Methods

Wenqin Shi,[1] Stephan W.H. Eijt,[1,a)] C.S. Suchand Sandeep,[2,b)] Laurens D.A. Siebbeles,[2] Arjan J. Houtepen,[2] Sachin Kinge,[3] Ekkes Brück,[1] Bernardo Barbiellini,[4] and Arun Bansil[4]

[1] *Department of Radiation, Radionuclides and Reactors, Faculty of Applied Sciences, Delft University of Technology, Mekelweg 15, NL-2629 JB, Delft, The Netherlands*

[2] *Department of Chemical Engineering, Faculty of Applied Sciences, Delft University of Technology, Julianalaan 136, NL-2628 BL, Delft, The Netherlands*

[3] *Toyota Motor Europe, Materials Research & Development, Hoge Wei 33, B-1930 Zaventem, Belgium*

[4] *Department of Physics, Northeastern University, 360 Huntington Avenue, Boston, MA 02155, USA*

---

a) Electronic mail: s.w.h.eijt@tudelft.nl
b) Present Address: *Institute of Physics and Astronomy, University of Potsdam, Karl-Liebknecht-Strasse 24-25, 14476 Potsdam, Germany*

**Abstract**

Positron Two Dimensional Angular Correlation of Annihilation Radiation (2D-ACAR) measurements reveal modifications of the electronic structure and composition at the surfaces of PbSe quantum dots (QDs), deposited as thin films, produced by various ligands containing either oxygen or nitrogen atoms. In particular, the 2D-ACAR measurements on thin films of colloidal PbSe QDs capped with oleic acid ligands yield an increased intensity in the electron momentum density (EMD) at high momenta compared to PbSe quantum dots capped with oleylamine. Moreover, the EMD of PbSe QDs is strongly affected by the small ethylediamine ligands, since these molecules lead to small distances between QDs and favor neck formation between near neighbor QDs, inducing electronic coupling between neighboring QDs. The high sensitivity to the presence of oxygen atoms at the surface can be also exploited to monitor the surface oxidation of PbSe QDs upon exposure to air. Our study clearly demonstrates that positron annihilation spectroscopy applied to thin films can probe surface transformations of colloidal semiconductor QDs embedded in functional layers.

Semiconductor quantum dots (QDs) present remarkable advantages as photovoltaic materials due to their tunable electronic structure. Physical properties of QDs such as photoluminescence (PL), charge carrier mobility, and multiple-exciton generation are greatly influenced by ligands used in the passivation of surfaces. The local environment of atoms on QD surfaces is intrinsically different from that of the core of QDs and varies with the type of ligand. The structural and electronic modifications due to surface ligands have been investigated via ab-initio modelling studies.[1-4] Experimental techniques such as X-ray absorption spectroscopy (XAS)[5] have been used to explore changes in electronic properties due to the nature of ligands. This is important since key properties such as photoconductivity and charge carrier mobility depend strongly on the type of ligand attached to the surface of the PbSe QDs.[6] Treating PbSe QD films with short ligand molecules such as ethylenediamine (EDA) leads to strong electronic coupling between the PbSe cores and results in high carrier mobility and photoconductivity.[6] A key issue in the development of solar cells based on PbSe QD absorber layers[7,8] is that these cells often show a strong, time-dependent degradation under ambient conditions.[9-11] Optical Absorption Spectroscopy (OAS) and X-ray Photoemission Spectroscopy (XPS) studies of PbSe QD films indicate that the dominant degradation process involves oxidation of the QDs through exposure to atmospheric oxygen. Strategies based on surface engineering, e.g involving the reaction of molecular chlorine with Se to form a protective thin $PbCl_x$ shell, are being developed to arrive at air-stable PbSe QD solids for applications as solar cells and field effect transistors.[12,13] Clearly, the development of methods to investigate and monitor surfaces of PbSe QDs embedded in a sub-surface layer in such devices is of crucial importance. In this regard, positron annihilation spectroscopy (PAS) offers unique advantages over XAS, OAS and XPS, since it combines a high sensitivity to selectively probe surfaces of nanoparticles[14-16] with an established capacity for depth-profiling of films in the range of ~10 nm to a few μm.[17-19] These unique merits are highly important in studies of photovoltaic devices, where charge carrier separation requires the formation of a p-n junction,[8] and PAS can probe the involved light absorbing and charge separation layers independently.

Here, we use the positron 2D-ACAR technique[14] to probe the electronic and compositional changes at the surfaces of PbSe QDs deposited as thin films. Previous positron lifetime experiments have shown that more than 90% of positrons implanted in a layer of PbSe QDs capped with Oleic Acid (OA) ligands trap at the surface of PbSe QDs,[16] where the positron is bound in a potential well behaving like a image potential at large distances (Figure 1).[20,21] Positrons implanted in the PbSe QD layer primarily lose their high initial kinetic energy in the PbSe cores of the QDs where they thermalize. Subsequently, they have a high probability of trapping in a surface state (Figure 1).[20,21] Therefore, the gamma rays produced by the annihilation of positrons trapped in surface states yields a way to probe the surface of PbSe QDs using 2D-ACAR, as the angular correlation of this annihilation radiation carries detailed information on the electron momentum density (EMD) sampled by the positron trapped in the surface state.[14]

Films with PbSe QDs using three different ligands, namely, Oleylamine (OLA), OA and EDA,[22] were studied in order to investigate the variation in electron-positron momentum density with the type of ligand end group (amine or carboxyl) and chain length (long or short). The two ligands OA and OLA are equally long, but differ in their attachment to the PbSe QDs. OA ligands attach to the surfaces of PbSe QDs by the formation of $Pb^{2+}(OA^-)_2$ complexes involving electrostatic interactions between lead and oxygen ions,[23,4,9] while the amine group of OLA only binds weakly to the surface of PbSe QDs, involving weak covalent interactions. Further, part of the QD surfaces might be passivated by chloride ions present in the synthesis.[24,25] EDA ligands are much shorter, and also interact only weakly to the surface of PbSe QDs. EDA treatment of PbSe QD films with OLA ligands mostly induces removal of the original ligands,[6] as was observed for films of PbSe QDs with OA ligands, while very little EDA was found on the surface after the treatment.[23] Our ACAR measurements on PbSe QD films with OA ligands reveal a stronger electron momentum density at high momenta compared to PbSe QDs with OLA ligands or treated with EDA. This effect is caused by oxygen atoms at the surface of QDs with OA ligands. This high sensitivity to oxygen atoms was further exploited to monitor the oxidation process of a layer of PbSe QDs with EDA ligands upon exposure to air.

PbSe QDs with average sizes ranging between 5 nm and 6 nm were synthesized by using oleylamine-based synthesis.[26] PbSe QDs with OA ligands and PbSe QDs treated with EDA were made by exchanging oleylamine with the other ligands. Hexane was used as a solvent for drop casting PbSe QDs with OA ligands and OLA ligands onto indium-tinoxide coated glass substrates. The deposition of the PbSe QDs treated with EDA was performed by dip coating. The PbSe QD films were examined by 2D-ACAR using the thin-film POSH-ACAR setup at the Reactor Institute Delft.[14-16] For comparison, a PbSe single crystal was studied by 2D-ACAR using a $^{22}$Na positron source. The 2D-ACAR distributions were obtained with positron implantation energies of 1 keV for PbSe QDs with EDA and 3.4 keV for PbSe QDs with OLA and OA ligands. These energies correspond to average implantation depths of ~7 nm and ~50 nm, respectively, coinciding to the middle of the deposited films as determined by positron Doppler broadening depth profiling.[22] The 2D-ACAR spectra consisting of $10^7$ counts were analysed using the ACAR2D program.[27] The distributions were isotropic due to the polycrystalline random orientation of the nanocrystals. Therefore, 1D-ACAR spectra were obtained by integrating the isotropic part of the 2D-ACAR spectra over one momentum direction.

Figure 2 presents the measured 1D-ACAR momentum distributions of PbSe QDs with OLA and OA ligands, and after EDA treatment of PbSe QDs with OLA ligands, in the form of ratio curves relative to bulk crystalline PbSe. The ratio curves show a peak at ~1 a.u. caused mostly by quantum confinement of the Se(4p) valence electron orbitals,[16] positron confinement[28,29] and other small contributions of atoms at the surface connected with the ligands.[16] The confinement of valence electrons and positrons for PbSe QDs capped with OA and OLA ligands leads to a small difference in the 1 a.u. peak area,[16,30] since the QDs have comparable sizes and the ligands have the same large chain length. In the case of EDA treated films, however, the confinement peak is significantly smaller. This pronounced difference can be attributed to the small QD-QD distances, which leads to electronic coupling of neighbouring QDs[6], resulting in a corresponding reduction of the Se(4p) valence electron confinement.[15] Furthermore, in ligand-exchange studies, it was demonstrated that short-chain diamines effectively strip lead oleate from (100) surfaces of the PbSe QDs.[23,31] The ligand replacement remains incomplete, particularly for the PbSe {100}-facets, which remain partially bare.[23] This leads to neck formation and epitaxially connection between PbSe QDs,[23] inducing electronic coupling between QDs. Partial depletion of QD surfaces and neck formation between QDs also occurs for EDA treatment of PbSe QDs capped with OLA ligands.[6] The electronic coupling induced by short QD-QD distances and neck formation results in more delocalized electron states with lower kinetic energy and momentum. This effect explains the narrowing of the EMD in momentum space, and a lower intensity of the confinement peak (Figure 2).

Independently, the high momentum region between 2 and 2.5 a.u. can be considered since the intensity here is primarily determined by the contributions of Pb atoms [Pb(5d) semi-core electrons][16] and contributions of ligand ad-atoms present at the surface. This region may also be influenced by positron annihilation with Se(3d) semi-core electrons.[16] The differences in the ratio curves in this momentum range reflect the various surface compositions probed by the trapped positrons. Previous calculations showed that the contribution of O(1s) electrons gradually becomes the most important factor in the observed electron-positron momentum density in the region p>2 a.u. for PbSe QDs with OA ligands.[16] Figure 2 shows that PbSe QDs with OA ligands exhibit a higher intensity in this momentum range than PbSe QDs with OLA ligands or PbSe QDs treated with EDA. OA and OLA share the same large $C_{17}H_{33}$ tail, and differ only in the –(C=O)O– and –(CH$_2$)HN– end groups attached strongly and weakly, respectively, to the surface of the PbSe QDs. The EMD of PbSe QDs observed by positron annihilation is thus clearly affected by the surface composition (involving the presence of oxygen versus nitrogen, and possibly chlorine), as is visible in the high momentum range. Indeed, calculations indicate that oxygen and chloride ions form attractive sites for positrons, as O$^-$ and Cl$^-$ form bound states with a positron, in contrast to the case of nitrogen.[32-34] One should note that, for a positron localized at the surface of the QD, the dominant Coulomb attraction comes mostly from an individual ion and the contribution from all other ions can be neglected as a first approximation as in the Boev-Arefiev model considered in Ref. [35].

Compared to the case of OLA ligands, PbSe QDs treated with EDA show a further reduced intensity in the high-momentum range, which could be attributed to annihilation at partially bare PbSe facets created during treatment with EDA.[23] Indeed, our previous study on PbSe QDs capped with OA ligands revealed a similar reduction in intensity for PbSe QDs with incompletely covered surfaces.[16] Confinement of valence electron orbitals with/without coupling between neighbouring QDs mainly influences the EMD at low momenta (Figure 2). This sensitivity demonstrates that thin film positron methods is a suitable characterization method to examine the surface composition and electronic structure of PbSe QD layers, which are key to understanding the resulting (opto-)electronic properties. Moreover, access to the depth range of up to ~1.5 μm below the outer surface of a PbSe QD layer in positron beam experiments enables one to perform full depth-profiling studies of innovative photovoltaic devices based on PbSe QD layers[13] as the involved layer thicknesses typically range from several hundred nanometers to micrometers.

In the second set of experiments, the sensitivity to detect oxygen present at the surface of PbSe QDs is further exploited. The local oxidation of the surfaces of PbSe QDs treated with EDA ligands can therefore be investigated. A rather thin PbSe QD film of about 2-3 monolayers deposited on a ~100 nm ITO layer on a glass substrate was used,[22] in order to prevent any depth dependence of the oxidation process as all PbSe QDs are uniformly exposed to oxygen of ambient air. Figure 3a shows that the first absorbance peak in the OAS spectra shifts towards shorter wavelength over a period of 3 months of oxidation under ambient conditions. The observed large blue shift of the absorption features points to a reduction in the effective PbSe core diameter from an initial value of 6.1 nm to 5.0 nm[11] caused by the formation of a thin oxidized shell. Figure 3b explicitly shows the gradual reduction of the effective PbSe core radius of the QDs as a function of exposure time in air, extracted from the wavelength of the first absorption peak.[36] The decrease of the effective radius of PbSe QD cores by 0.3 to 0.5 nm over a period of 1 to 2 months as revealed by the OAS spectra is close to what was observed in previous studies on ~6 nm diameter PbSe QDs stored in solution under ambient conditions and application of vigorous stirring.[11]

Figure 4a shows the corresponding evolution of 1D-ACAR ratio curves of PbSe QDs treated with EDA, for different periods of air exposure. The oxidation of PbSe QDs after 24 days leads to an increase of the confinement peak near 1 a.u. due to the decrease in effective size of the core PbSe QDs by the oxidation, which enhances the confinement of the Se(4p) valence electrons and also reduces the electronic coupling between QDs. Interestingly, the increased fraction of oxygen atoms present at the surfaces of PbSe QDs by surface oxidation yields a pronounced increase in intensity for $p > 2$ a.u. due to the contribution of O(1s) electrons to the momentum density. Notably, the strongest increase in the momentum density in the range of 2-2.5 a.u. is seen already after the first period of exposure in air of 24 days (Figure 4), while the oxide shell continues to grow afterwards, as revealed by the continued shift towards shorter wavelengths of the OAS spectra (Figure 3). The intensity ratio to bulk PbSe in the momentum range of 2-2.5 a.u. (Figure 4b), in contrast, indicates a saturation in the high momentum intensity for the longer air exposure times of 61 and 90 days. This can be understood since the positron mostly probes the outermost atomic layers of the QD, *i.e.* the outer surface region of the oxidized shell, likely consisting of PbO, SeO$_2$ and PbSeO$_3$ formed at the surface of the QD.[10] Small differences in intensity at high momenta over 24-90 days can be caused by a variation in binding energy and distance of the positron from the surface, since the image potential will reduce in strength due to the more insulating character of the surface with the growth of an oxide layer.[37] The transformation of the surface by the oxidation process thus seems to saturate sooner than the continued growth of the oxidized shell below the surface, as monitored by the first absorption peak in the OAS which provides a sensitive measure of the size of the non-oxidized PbSe cores of the QDs. Decrease in the effective PbSe core size revealed by the OAS is consistent with the observed (small) increase in the confinement peak in the 1D-ACAR ratio curves, while the intensity of the ratio curves for p > 2 a.u. provides complementary information on the local oxygen fraction formed at the outer surface of the PbSe QDs during the oxidation process. These results demonstrate that positron methods can be sensitively applied to probe the oxidation at the surfaces of colloidal QDs, and indicate more generally that surface chemical transformations can be monitored. A more quantitative determination of the surface chemical composition requires details of shape of the positron wave function at the surface of the QD and its overlap with the (local) electronic orbitals, and the development of appropriate ab-initio methods for this purpose is needed.

In summary, our study demonstrates that positrons are effective in probing the composition of the surfaces of reduced dimensional systems such as the PbSe QDs. By enabling the advanced characterization of the attached ligand molecules, our method can facilitate the optimization of efficient charge carrier transport in optoelectronic devices by introducing descriptors based on positron annihilation characteristics. Our study further demonstrates that positron methods[38] can be used to sensitively monitor oxidation processes at the surfaces of colloidal nanocrystals. Thin film positron methods thus hold promise as a surface characterization technique for colloidal semiconductor QDs in functional layers used to develop (opto-)electronic devices including thin film solar cells, light-emitting diodes and field effect transistors. Positron techniques will thus provide key insights into chemical transformations at the surfaces of QDs and aid their development by surface chemical engineering strategies of innovative core-shell structures required for stabilization against oxidation under ambient conditions. Finally, our study shows that the electron momentum density of PbSe QDs is strongly affected by electronic coupling between the neighbouring QDs induced by the use of small EDA molecules, related to short interparticle distances and induced neck formation between neighboring QDs. Such a coupling is very important to achieve favorable charge carrier mobility and enhanced photoconductivity of PbSe QD layers for prospective application in next generation photovoltaic devices. The depth-profiling capability of the positron method can provide a potential *in-situ* tool for probing light absorbing and charge separation layers as well as the p-n junction interface region independently, which is important for realizing substantial improvement in solar cell efficiencies.

The work at Delft University of Technology was supported by a China Scholarship Council (CSC) grant, and ADEM, A green Deal in Energy Materials of the Ministry of Economic Affairs of The Netherlands (www.adem-innovationlab.nl). C.S.S.S. acknowledges funding by Toyota Motor Europe. The work at Northeastern University was supported by the US Department of Energy (DOE), Office of Science, Basic Energy Sciences grant number DE-FG02-07ER46352, and benefited from Northeastern University's Advanced Scientific Computation Center (ASCC) and the NERSC supercomputing center through DOE grant number DE-AC02-05CH11231. The authors would like to thank Martijn de Boer for technical support in the positron experiments.

**Figures**

Figure 1.

(Color online) Schematic density $|\psi_+|^2$ of the positron surface state of the quantum dot, and schematic illustration of a $Pb^{2+}(OA^-)_2$ complex at the surface.

Figure 2.

(Color online) The ratio curve of 1D-ACAR momentum distributions for PbSe QDs with various surface compositions: oleic acid (blue); oleylamine (red); and, EDA (magenta). All curves are normalized to the directionally averaged 1D-ACAR distribution of bulk PbSe. (a.u. = atomic units).

Figure 3.

(Color online) a) Evolution of absorption spectra of a PbSe QD film treated with EDA upon exposure to air at room temperature; b) Evolution of the effective PbSe core size of the QDs. Solid line serves as a guide-to-the-eye.

Figure 4.

(Color online) a) Evolution of 1D-ACAR momentum distribution for PbSe QDs treated with EDA with exposure time in air, presented as ratio curves relative to the directionally averaged 1D-ACAR distribution of bulk PbSe; b) Normalized-intensity ratio to bulk PbSe in the momentum region between 2 a.u. and 2.5 a.u. Solid line is a guide-to-the-eye.

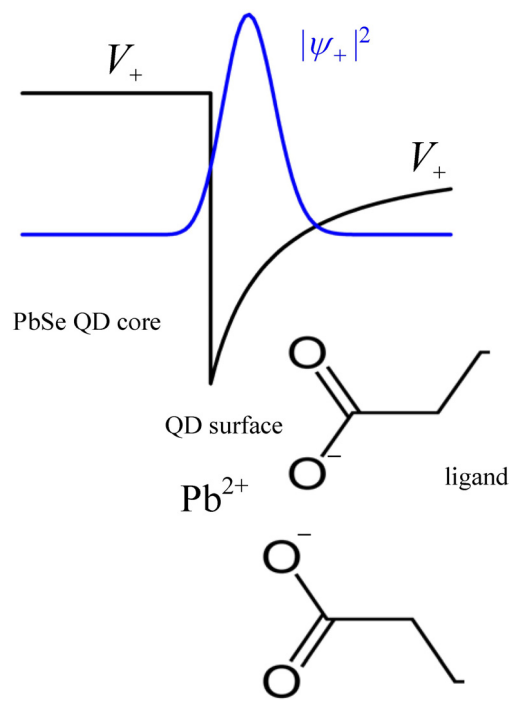

Figure 1.

(Color online) Schematic density $|\psi_+|^2$ of the positron surface state of the quantum dot, and schematic illustration of a $Pb^{2+}(OA^-)_2$ complex at the surface.

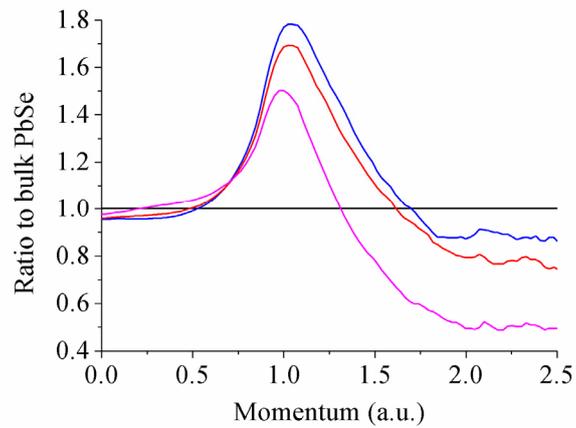

Figure 2.

(Color online) The ratio curve of 1D-ACAR momentum distributions for PbSe QDs with various surface compositions: oleic acid (blue); oleylamine (red); and, EDA (magenta). All curves are normalized to the directionally averaged 1D-ACAR distribution of bulk PbSe. (a.u. = atomic units).

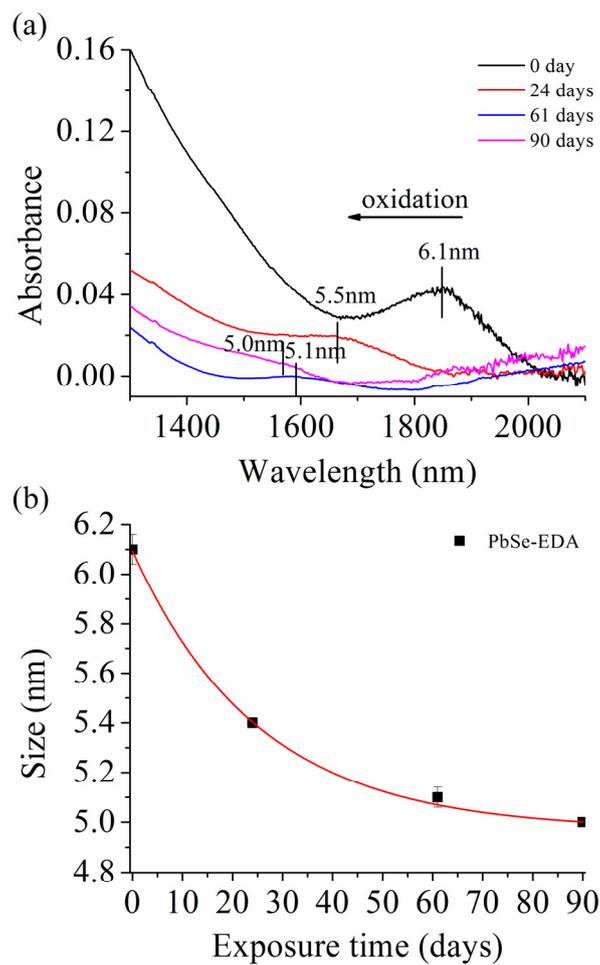

Figure 3.

(Color online) a) Evolution of absorption spectra of a PbSe QD film treated with EDA upon exposure to air at room temperature; b) Evolution of the effective PbSe core size of the QDs. Solid line serves as a guide-to-the-eye.

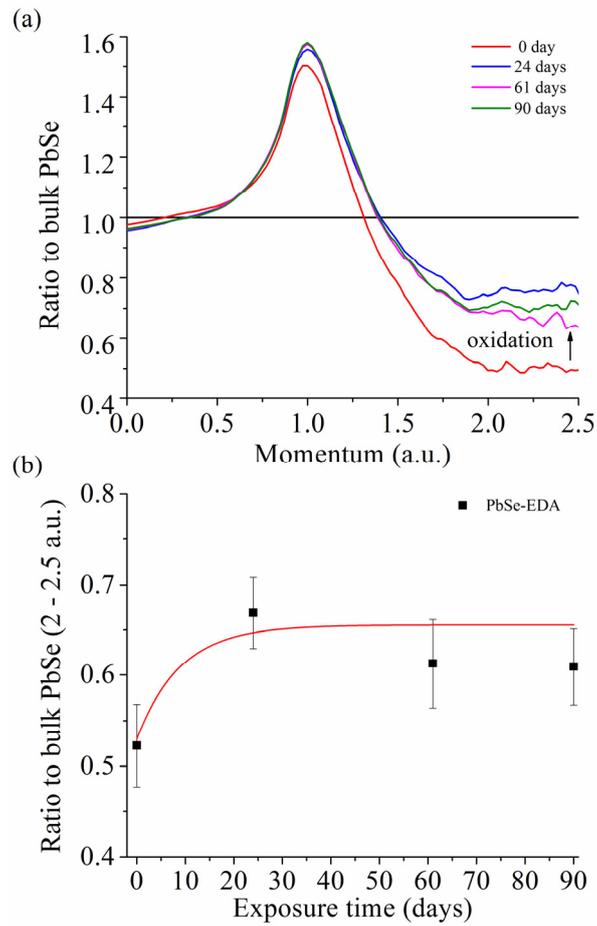

Figure 4.

(Color online) a) Evolution of 1D-ACAR momentum distribution for PbSe QDs treated with EDA with exposure time in air, presented as ratio curves relative to the directionally averaged 1D-ACAR distribution of bulk PbSe; b) Normalized-intensity ratio to bulk PbSe in the momentum region between 2 a.u. and 2.5 a.u. Solid line is a guide-to-the-eye.

# Supplemental Material

## Ligand-Surface Interactions and Surface Oxidation of Colloidal PbSe Quantum Dots Revealed by Thin-film Positron Annihilation Methods

*Wenqin Shi, Stephan W.H. Eijt,\* C.S. Suchand Sandeep, Laurens D.A. Siebbeles, Arjan J. Houtepen, Sachin Kinge, Ekkes Brück, Bernardo Barbiellini, and Arun Bansil*

XRD and OAS spectra of PbSe QD films

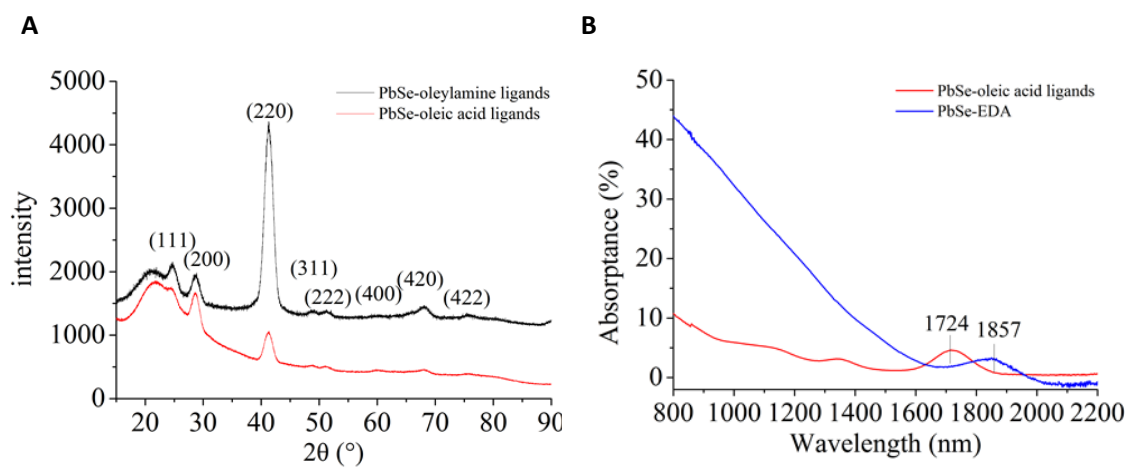

**Figure S1.** A) XRD spectra of PbSe QDs with oleylamine ligands and oleic acid ligands; B) OAS spectra of PbSe QDs with oleic acid ligands and PbSe QDs treated with EDA.

**Table S1:** sizes of PbSe QDs extracted from XRD and OAS.

| Sample | Average size(nm) | |
|---|---|---|
| | XRD | OAS |
| PbSe-oleic acid ligands | 6.4 | 5.6 |
| PbSe-oleylamine ligands | 6.1 | - |
| PbSe-EDA | - | 6.1 |

In Figure S1, XRD and OAS spectra are presented, which were used to estimate the average sizes of PbSe QDs listed in Table S1. The average as-deposited nanocrystal size of the oleic acid- and oleylamine-capped PbSe QDs estimated from the peak broadening of XRD spectra is 6.4 nm and 6.1 nm, respectively. From the first exciton absorption peak, the extracted sizes of the PbSe QDs with oleic acid ligands and of the PbSe QDs treated with EDA are 5.6 nm and 6.1 nm. The XRD patterns show that the PbSe QDs are in the rock salt phase.

Doppler depth profiles of PbSe QD film treated with EDA used in the oxidation study

The positron Doppler broadening of annihilation radiation with 511 keV was measured using positrons with a tunable kinetic energy in the range of 0-25 keV. The S and W parameters were determined using momentum windows for S and W of $|p| < 3.0 \times 10^{-3}$ $m_0c$ and $8.2 \times 10^{-3}$ $m_0c < |p| < 23.4 \times 10^{-3}$ $m_0c$, respectively, with $m_0$ and c respectively the electron rest mass and the light velocity. The S parameter is a measure for positron annihilation with valence electrons, which provides sensitivity to the electronic structure and the presence of vacancies. The W parameter is a measure of annihilation with semi-core electrons. which provides chemical sensitivity to the positron trapping site. The S and W depth profiles were analyzed using the VEPFIT program package.[1] The detailed procedures for the positron Doppler broadening depth profiling method were described in Ref. 2.

A thickness of the PbSe QD film with EDA of about 15 nm (2-3 monolayers of PbSe QDs) was extracted from VEPFIT analysis of the positron Doppler broadening depth profiles shown in Figure S2. The fitted parameters are listed in Table S2 and S3.

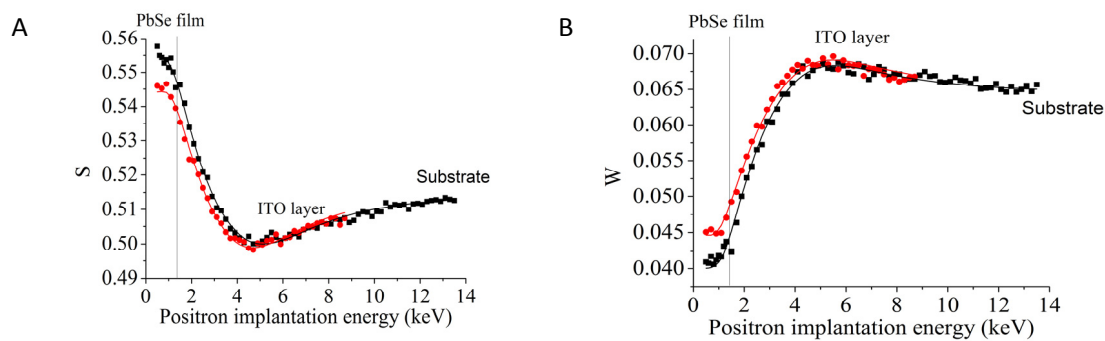

**Figure S2.** Positron Doppler broadening S-parameter (W-parameter) depth profiles of PbSe QD layer on an ITO-coated glass substrate before (black squares) and after oxidation (red circles).

Table S2: VEPFIT analysis results for a PbSe QD film treated with EDA.

| layer | Density (g/cm³) | Diffusion length (nm) | Layer width (nm) | S parameter | W parameter |
|---|---|---|---|---|---|
| PbSe layer | 5.67* | 5 | 15 | 0.5558 (0.0005) | 0.039 (0.001) |
| ITO | 7.1* | 16* | 109 | 0.4902 (0.0003) | 0.073 (0.001) |
| glass | 2.53* | 29* | - | 0.5149 (0.0003) | 0.064 (0.001) |

Table S3: VEPFIT analysis results for an oxidized PbSe QD film treated with EDA.

| layer | Density (g/cm$^3$) | Diffusion length (nm) | Layer width (nm) | S parameter | W parameter |
|---|---|---|---|---|---|
| PbSe layer | 5.67* | 1 | 12 | 0.5443 (0.0004) | 0.045 (0.001) |
| ITO | 7.1* | 16* | 93 | 0.4871 (0.0004) | 0.074 (0.001) |
| glass | 2.53* | 29* | - | 0.5149 (0.0003) | 0.064 (0.001) |

* fixed

Positron 2D-ACAR

The PbSe quantum dot films were examined by two-dimensional angular correlation of annihilation radiation (2D-ACAR)[3,4] using the thin-film POSH-ACAR setup at the Reactor Institute Delft. For comparison, a PbSe single crystal was studied by 2D-ACAR using a $^{22}$Na positron source. The 2D-ACAR technique measures the two-dimensional projection of the momentum distribution of the annihilated e$^-$-e$^+$ pair, which is determined by the square of the overlap integral of the electron wavefunction ($\psi^-$) and positron wave function ($\psi^+$), summed over all occupied electron states. The momentum p of the annihilated e$^-$-e$^+$ pair is reflected in the small deviation from collinearity of the emitted quanta.[3,4] Since the momentum of the thermalized positron is reduced to almost zero before the annihilation, p essentially represents the momentum of the electron. The thicknesses of the deposited QD films were examined by positron Doppler broadening depth profiling measurements (Figure S2) in the POSH-ACAR setup by using a Ge detector. These depth profiles were used to select the positron implantation energies used subsequently in the 2D-ACAR measurements. The 2D-ACAR distributions were obtained with positron implantation energies of 1 keV for PbSe QDs with EDA and 3.4 keV for PbSe QDs with oleylamine and oleic acid ligands, i.e., with average implantation depths of ~7 nm and ~50 nm, respectively. The momentum resolution was 1.4×1.4 (10$^{-3}$ m$_0$c)$^2$, where m$_0$ is the electron rest mass and c the velocity of light.[3,4] The 2D-ACAR spectra, each consisting of about 10$^7$ collected annihilation events, were analysed using the ACAR2D program.[5] 2D-ACAR distributions of the PbSe QD layers were isotropic due to the polycrystalline random orientation of the nanocrystals. Therefore, 1D-ACAR spectra were obtained by integrating the isotropic part of the 2D-ACAR spectra over one momentum direction.

## Ligand molecules

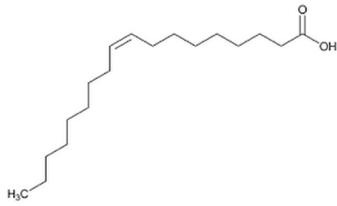
(a) Oleic acid (OA)

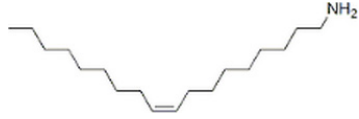
(b) Oleylamine (OLA)

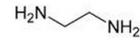
(c) Ethylenediamine (EDA)

| Ligand type | Ligand name | Molecular structure | Interparticle space | Ligand end group |
|---|---|---|---|---|
| Long-chain molecules | Oleylamine, Oleic acid | PbSe, Cn-tail n=18, □=COOH, NH$_2$ | >1.5nm | -(CH$_2$)NH$_2$ -(C=O)OH |
| Short-chain molecules | EDA | PbSe, Cn-tail n=2, □=NH$_2$ | <0.6nm | -(CH$_2$)NH$_2$ |

## Oxidation process

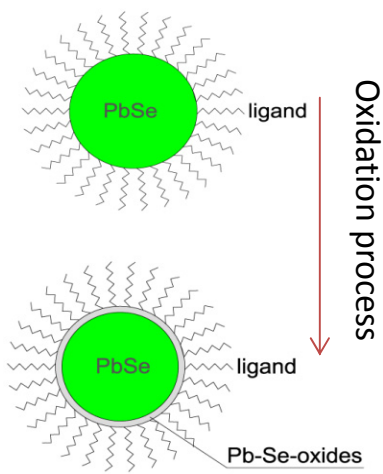